\documentclass{aip-cp}

\usepackage[numbers]{natbib}
\usepackage{rotating}
\usepackage{graphicx}

\newcommand{\ba}{\begin{eqnarray}}
\newcommand{\ea}{\end{eqnarray}}
\newcommand{\ban}{\begin{eqnarray*}}
\newcommand{\ean}{\end{eqnarray*}}
\newcommand{\bsub}{\begin{subequations}}
\newcommand{\esub}{\end{subequations}}

\def\ket#1{|#1\rangle}
\def\bra#1{\langle#1|}	
\def\bsu3{\overline{{\rm SU(3)}}}
\def\blam{\bar{\lambda}}
\def\bmu{\bar{\mu}}

\begin{document}

\title{Dynamical Symmetries and Beyond: Lessons and Advances}

\author{A. Leviatan}
\eaddress[url]{http://www.phys.huji.ac.il/$\sim$ami/}

\affil{Racah Institute of Physics, The Hebrew University, 
Jerusalem 91904, Israel}
\corresp{Corresponding author: ami@phys.huji.ac.il}

\maketitle

\begin{abstract}
  A central theme in Iachello's quest for understanding simple ordered
  patterns in complex quantum systems, is the concept of dynamical symmetry.  
  Relying on his seminal contributions, we present
  further generalization of this notion to that of partial dynamical
  symmetry (PDS), for which solvability and good quantum numbers are
  maintained by only a subset of states. Hamiltonians with a~single PDS
  and multiple PDSs are constructed explicitly and their relevance to
  nuclear structure is discussed.
\end{abstract}

\section{INTRODUCTION}
The concept of dynamical symmetry (DS)
is now widely recognized to play a pivotal role
in our understanding of dynamical systems.
In particular, it had a major impact
on developments in nuclear~\cite{ibm,ibfm}, 
molecular~\cite{vibron} and hadronic physics~\cite{BIL94}, 
pioneered by F.~Iachello and his colleagues.
Its basic paradigm is to write the Hamiltonian of the system,
\ba
\hat{H}_{\rm DS} = \sum_{G} a_{G}\,\hat{C}[G]\;\;\;\qquad\qquad
E_{\rm DS}(\lambda_{\rm dyn},\, \lambda_1,\,\lambda_2,\,\ldots,
\,\lambda_{\rm sym}) ~,
\label{H-DS}
\ea
in terms of the Casimir operators, $\hat{C}[G]$, of a chain of
nested algebras~\cite{LieAlg},
\ba
G_{\rm dyn} \supset G_1 \supset G_2 \supset \dots \supset G_{\rm sym} 
\;\;\;\qquad
\ket{\lambda_{\rm dyn},\, \lambda_1,\,\lambda_2,\,
\ldots,\,\lambda_{\rm sym}} ~.\qquad 
\label{DS-chain}
\ea
In such a case, the spectrum is completely solvable,
the eigenstates and energies, $E_{\rm DS}$, are labeled by quantum numbers 
$(\lambda_{\rm dyn},\, \lambda_1,\,\lambda_2,\,\ldots,\,\lambda_{\rm sym})$, 
which are the labels of 
irreducible representations (irreps) of the algebras in the chain.
In Eq.~(\ref{DS-chain}), $G_{\rm dyn}$ is the dynamical (spectrum generating) 
algebra of the system such that operators of all physical observables can 
be written in terms of its generators
and $G_{\rm sym}$ is the symmetry algebra.
A given $G_{\rm dyn}$ can encompass several DS chains, each providing 
characteristic analytic expressions for observables and definite 
selection rules.

A notable example of such algebraic construction
is the interacting boson model 
(IBM)~\cite{Arima75,Arima76,Arima78,Arima79},
describing low-lying quadrupole 
collective states in nuclei in terms of $N$ monopole $(s)$ and
quadrupole $(d)$ bosons, representing valence nucleon pairs.
The model is based on a unitary spectrum generating algebra
$G_{\rm dyn}\!=\!{\rm U(6)}$ and an orthogonal (angular-momentum) symmetry
algebra $G_{\rm sym}\!=\!{\rm SO(3)}$. The Hamiltonian
is expanded in the elements of U(6),
and consists of Hermitian, rotational-scalar interactions 
which conserve the total number of $s$- and $d$- bosons, 
$\hat N \!=\! \hat{n}_s + \hat{n}_d \!=\! 
s^{\dagger}s + \sum_{m}d^{\dagger}_{m}d_{m}$. 
The solvable limits of the IBM correspond to the following DS chains,
\ba
&&\hspace{-0.5cm}
{\rm U(6)\supset U(5)\supset SO(5)\supset SO(3)}
\;\,\quad
\ket{N,\, n_d,\,\tau,\,n_{\Delta},\,L} \quad\quad 
E_{\rm DS} = \epsilon_d\, n_d + A\,n_d(n_d+4)
+ B\,\tau(\tau+3) + C\,L(L+1) ~,\qquad
\label{U5-ds}
\\
&&\hspace{-0.5cm}
{\rm U(6)\supset SU(3)\supset SO(3)}
\;\;\qquad\quad\quad
\ket{N,\, (\lambda,\mu),\,K,\, L} \quad\quad
E_{\rm DS} = A(\lambda^2 +\mu^2 +\lambda\mu + 3\lambda + 3\mu)
+ C\,L(L+1) ~,\qquad\quad
\label{SU3-ds}
\\
&&\hspace{-0.5cm}
{\rm U(6)\supset \bsu3\supset SO(3)}
\;\;\qquad\quad\quad
  \ket{N,\, (\blam,\bmu),\,\bar{K},\, L} \quad\quad
E_{\rm DS} = A(\blam^2 +\bmu^2 +\blam\bmu + 3\blam + 3\bmu)
+ C\,L(L+1) ~,\quad\quad
\label{SU3bar-ds}
\\
&&\hspace{-0.5cm}
  {\rm U(6)\supset SO(6)\supset SO(5)\supset SO(3)}
\quad
\ket{N,\, \sigma,\,\tau,\,n_{\Delta},\, L} \quad\quad\;
E_{\rm DS} = A\,\sigma(\sigma+4)
+ B\,\tau(\tau+3) + C\,L(L+1) ~.\quad\quad
\label{SO6-ds}
\ea
Here $N,n_d,(\lambda,\mu),(\blam,\bmu),\sigma,\tau,L$, 
label the relevant irreps of 
U(6), U(5), SU(3), $\bsu3$, SO(6), SO(5), SO(3), 
respectively, and $n_{\Delta},K,\bar{K}$ are multiplicity labels. 
Each chain provides a complete basis whose members are eigenstates of the
Casimir operators in the chain with eigenvalues listed above.
The resulting spectra of these DS chains
with leading sub-algebras ${\rm G_1}$: 
U(5), SU(3), ${\rm\overline{SU(3)}}$ and SO(6), 
resemble known paradigms of nuclear collective 
structure: spherical vibrator, prolate-, oblate- and $\gamma$-soft
deformed rotors, respectively.
Electromagnetic moments and rates can be calculated 
with transition operators of appropriate rank. For example,
the one-body $E2$ operator reads $\hat{T}(E2)=$
$e_{B}[ d^{\dag}s + s^{\dag}\tilde{d} + \chi\,(d^{\dag}\tilde{d})^{(2)}]$,
where $\tilde{d}_{m} = (-1)^{m}d_{-m}$, and standard notation of 
angular momentum coupling is used.

A geometric visualization of the IBM is obtained by an energy surface,
\ba
E_{N}(\beta,\gamma) &=&
\bra{\beta,\gamma; N} \hat{H} \ket{\beta,\gamma ; N} ~,
\label{enesurf}
\ea 
defined by the expectation value of the Hamiltonian in a coherent 
(intrinsic) state~\cite{gino80,diep80}.
Here $(\beta,\gamma)$ are
quadrupole shape parameters whose values, $(\beta_{\rm eq},\gamma_{\rm eq})$, 
at the global minimum of $E_{N}(\beta,\gamma)$ define the equilibrium 
shape for a given Hamiltonian. 
The shape can be spherical $(\beta \!=\!0)$ or 
deformed $(\beta \!>\!0)$ with $\gamma \!=\!0$ (prolate), 
$\gamma \!=\!\pi/3$ (oblate), 
$0 \!<\! \gamma \!<\! \pi/3$ (triaxial) or $\gamma$-independent.
The equilibrium deformations associated with the 
DS limits of Eqs.~(3)-(6), conform with their geometric interpretation,
and are given by 
$\beta_{\rm eq}\!=\!0$ for U(5), 
$(\beta_{\rm eq} \!=\!\sqrt{2},\gamma_{\rm eq}\!=\!0)$ for SU(3), 
$(\beta_{\rm eq} \!=\!\sqrt{2},\gamma_{\rm eq}\!=\!\pi/3)$ for $\bsu3$, 
and $(\beta_{\rm eq}\!=\!1,\gamma_{\rm eq}\,\,{\rm arbitrary})$ for SO(6). 
The DS Hamiltonians support a single minimum in their 
energy surface, hence serve as benchmarks for the dynamics of a single 
quadrupole shape.

Since its introduction, the IBM has been the subject of many investigations,
becoming a standard model for the description of atomic nuclei.
An important lesson from these extensive studies is the observation that,
although a dynamical symmetry
provides considerable insights, in most applications to realistic systems,
the predictions of an exact DS are rarely fulfilled and one is compelled
to break it.
More often one finds that the assumed symmetry is not obeyed uniformly,
{\it i.e.}, is fulfilled by some of the states but not by others.
The need to address such situations, but still preserve
important symmetry remnants, has motivated
the introduction of partial dynamical symmetry (PDS)~\cite{Lev11}.
The essential idea is to relax the stringent conditions of
{\it complete} solvability so that only part of the eigenspectrum retains 
analyticity and/or good quantum numbers.
The novel notion of PDS and its implications to nuclear structure
are the subject matter of the present contribution.

\section{PARTIAL DYNAMICAL SYMMETRY AND NUCLEAR SPECTROSCOPY}

A partial dynamical symmetry (PDS) corresponds to a particular
symmetry-breaking for which the virtues of a dynamical symmetry (DS),
namely, solvability and good quantum numbers, are fulfilled by only
a subset of states. The IBM, with its rich algebraic structure, provides
a convenient framework for realizing the PDS notion and testing its
predictions. Consider one of the DS chains of the IBM, Eqs.~(3)-(6),
\ba
{\rm U(6)\supset G_1\supset G_2\supset \ldots \supset SO(3)} &&\;\;\quad
\ket{N,\, \lambda_1,\,\lambda_2,\,\ldots,\,L} ~, 
\label{u6-ds}
\ea  
with leading sub-algebra $G_1$ and basis
$\ket{N,\lambda_1,\lambda_2,\ldots,L}$.
The algorithm for constructing Hamiltonians with PDS, associated
with the reduction~(\ref{u6-ds}), is based on identifying $n$-particle
annihilation operators $\hat{T}_{\alpha}$ which
satisfy~\cite{AL92,RamLevVan09},
\ba
\hat{T}_{\alpha}
\ket{N,\lambda_1\!=\!\Lambda_0,\lambda_2,\ldots,L} = 0 ~.
\label{Talpha}
\ea
The set of states in Eq.~(\ref{Talpha}) are basis states of a particular
$G_1$-irrep, $\lambda_1=\Lambda_0$, with good $G_1$ symmetry,
and are specified by the quantum numbers of the algebras in the
chain~(\ref{u6-ds}).
These states may span the entire or part of the indicated irrep.
Condition~(\ref{Talpha}) ensures that they 
are zero-energy eigenstates of the following normal-ordered Hamiltonian,
\ba
\hat{H} &=& 
\sum_{\alpha,\beta}u_{\alpha\beta}\hat{T}^{\dag}_{\alpha}\hat{T}_{\beta} ~.
\label{H-int}
\ea
$\hat{H}$ itself, however, need not be invariant 
under $G_1$ and, therefore, has partial-$G_1$ symmetry. 
The degeneracy of the above set of states 
is lifted, without affecting their wave functions,
by adding to $\hat{H}$ the following Hamiltonian
\ba
\hat{H}_c &=& \sum_{G_i\subset G_1} a_{G_i}\hat{C}[G_i] ~,
\label{H-col}
\ea
composed of the Casimir operators of the
sub-algebras of $G_1$ in the chain~(\ref{u6-ds}).
The states $\ket{N,\lambda_1\!=\!\Lambda_0,\lambda_2,\ldots,L}$
remain solvable eigenstates of the complete Hamiltonian 
\ba
\hat{H}_{\rm PDS} &=& \hat{H} + \hat{H}_c ~,
\label{H-PDS}
\ea
which, by definition, has $G_1$-PDS.
In the nuclear physics terminology, the decomposition of Eq.~(\ref{H-PDS})
is referred to as a resolution of the Hamiltonian into intrinsic
($\hat{H}$) and collective ($\hat{H}_c$) parts~\cite{kirlev85,lev87}.
The former determines the energy surface~(\ref{enesurf})
and band-structure, while the latter
determines the in-band rotational splitting.
Since $\hat{H}$ is related to the Casimir
operator of $G_1$ for a specific choice of parameters,
the PDS-Hamiltonian of Eq.~(\ref{H-PDS}) can be also transcribed in the form
\ba
\hat{H}_{\rm PDS} &=& \hat{H}_{\rm DS} + \hat{V}_0  ~,
\label{H-PDSV0}
\ea
where $\hat{H}_{\rm DS}$ is the DS
Hamiltonian, Eq.~(\ref{H-DS}), for the chain (\ref{u6-ds}) and
$\hat{V}_0$ satisfies
$\hat{V}_0\ket{N,\lambda_1\!=\!\Lambda_0,\lambda_2,\ldots,L}=0$.
In what follows, we present explicit
PDS Hamiltonians associated with
the DS chains of the IBM, and show their relevance to nuclear spectroscopy.

The spectrum corresponding to the SU(3)-DS chain of Eq.~(\ref{SU3-ds}),
resembles that of a prolate-deformed roto-vibrator.
The eigenstates are arranged in SU(3) $(\lambda,\mu)$-multiplets,
forming rotational $K$-bands with characteristic $L(L+1)$ splitting.
The label $K$ corresponds geometrically to the projection of the angular
momentum on the symmetry axis. The lowest SU(3) irrep $(2N,0)$ contains
the ground band g$(K\!=\!0)$, and the irrep $(2N-4,2)$ contains
both the $\beta(K\!=\!0)$ and $\gamma(K\!=\!2)$ bands.
\begin{figure}[t]
\centerline{
\fbox{\includegraphics[width=0.54\linewidth]{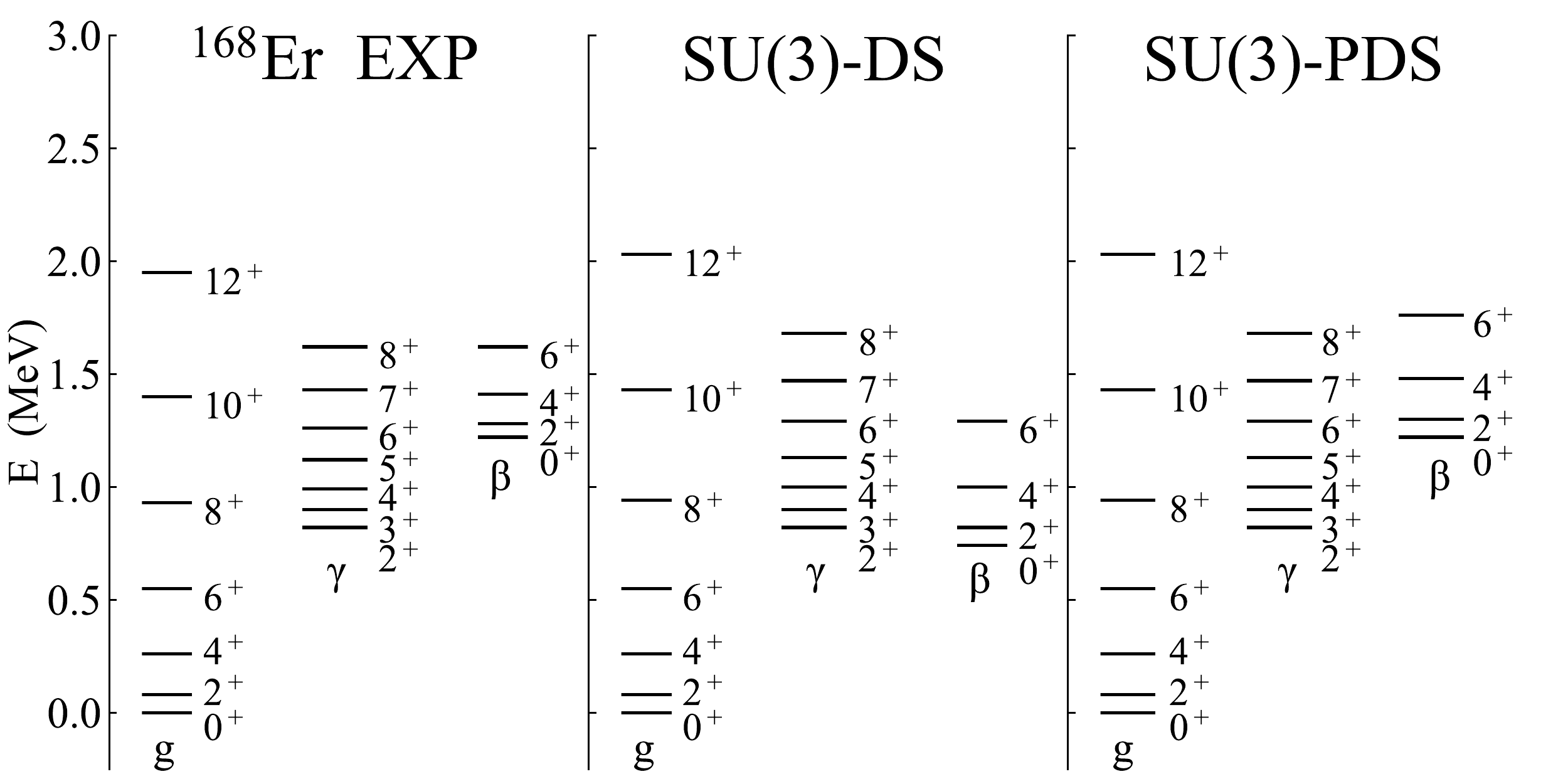}}
\hspace{0.06cm}%
\includegraphics[width=0.45\linewidth]{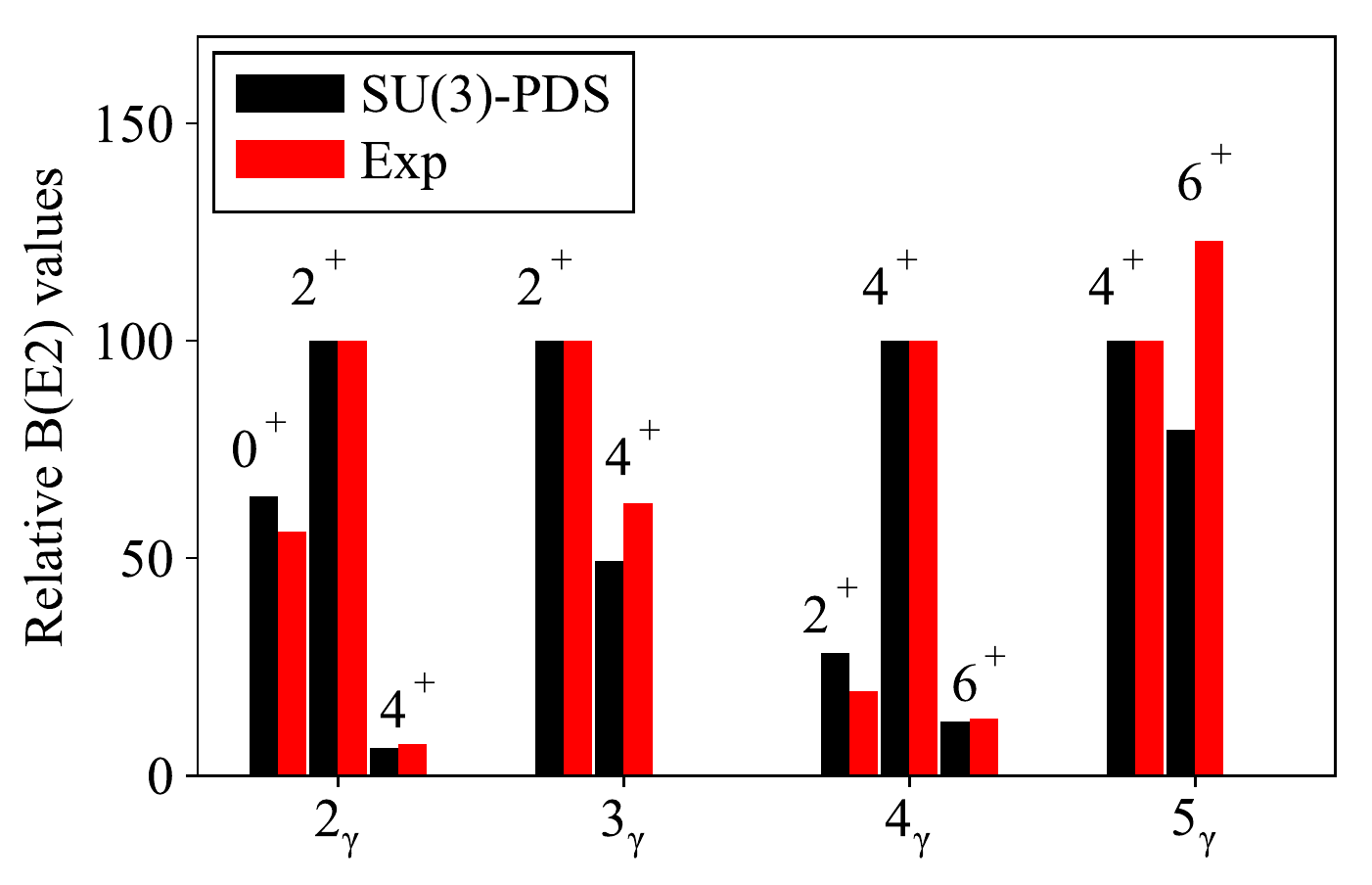}}
  \caption{
\small
Left panels: observed spectrum of $^{168}$Er compared with SU(3)-DS and
SU(3)-PDS calculations. The latter employs $\hat{H}_{\rm PDS}$
of Eq.~(\ref{HSU3-PDS})
with $h_0\!=\!8,\,h_2\!=\!4,\,C\!=\!13$ keV
and $N\!=\!16$. Right panel:
comparison of the PDS parameter-free predictions with the data on the
relative $B(E2; L_{\gamma}\to L)$ values for $\gamma\to g$ $E2$ transitions
in $^{168}$Er. Adapted from ~\cite{lev96,Casten14}.
\label{figEr168}}
\end{figure}

The construction of Hamiltonians with SU(3)-PDS follows the general
algorithm. The two-boson pair operators 
\ba
P^{\dagger}_{0} &=& d^{\dagger}\cdot d^{\dagger} - 2(s^{\dagger})^2 ~,
\label{P0}
\\
P^{\dagger}_{2m} &=& 2d^{\dagger}_{m}s^{\dagger} + 
\sqrt{7}\, (d^{\dagger}\,d^{\dagger})^{(2)}_{m} ~,
\label{P2}
\ea
are $(\lambda,\mu)=(0,2)$ tensors of SU(3), 
and annihilate all $L$-states of the SU(3) ground band
irrep $(\lambda,\mu)=(2N,0)$,
\ba
P_{0}\,\ket{N,\, (\lambda,\mu)\!=\!(2N,0),\,K\!=\!0,\, L} &=& 0 ~,
\qquad\quad L=0,2,4\ldots,2N
\nonumber\\
P_{2m}\,\ket{N,\, (\lambda,\mu)\!=\!(2N,0),\,K\!=\!0,\, L} &=& 0 ~.
\label{P0P2}
\ea 
In addition, $P_0$ satisfies 
\ba
P_{0}\,\ket{N,\, (\lambda,\mu)\!=\!(2N-4k,2k),\,K\!=\!2k,\, L} = 0 ~,\;\;\;\;
L=K,K+1,\ldots, (2N-2k)~.
\label{P0k}
\ea
For $k> 0$, the $L$-states of Eq.~(\ref{P0k}) 
span only part of the SU(3) irreps 
$(\lambda,\mu)=(2N-4k,2k)$ and form the rotational members of excited 
$\gamma^{k}(K=2k)$ bands.
$P_0$ and $P_{2m}$ correspond to the operators 
$\hat{T}_{\alpha}$ of Eq.~(\ref{Talpha}).
The intrinsic Hamiltonian reads
$\hat{H} = h_{0}\,P^{\dagger}_{0} P_0
+ h_{2}\,P^{\dagger}_{2}\cdot \tilde{P}_{2}$,
where $\tilde{P}_{2m} = (-)^{m}P_{2,-m}$
and the centered dot denotes a scalar product.
The collective Hamiltonian, $\hat{H}_c = C\,\hat{C}_{2}[\rm SO(3)]$,
is composed of the quadratic Casimir operator of SO(3),
with eigenvalues $L(L+1)$. The complete Hamiltonian with SU(3)-PDS
has the form as in Eqs.~(\ref{H-PDS}) and (\ref{H-PDSV0}),
and is given by~\cite{lev96}, 
\ba
\hat{H}_{\rm PDS} =
h_0P^{\dagger}_{0}P_{0} + h_2P^{\dagger}_{2}\cdot \tilde{P}_{2}
+ C\,\hat{C}_{2}[\rm SO(3)]
= \hat{H}_{\rm SU(3)-DS} + \eta_0\,P^{\dagger}_{0}P_{0} ~.
\label{HSU3-PDS}
\ea
The second equality in Eq.~(\ref{HSU3-PDS}) follows from the fact that for
$h_2\!=\!h_0$, the combination
$P^{\dagger}_{0}P_{0} + P^{\dagger}_{2}\cdot \tilde{P}_{2}  = 
-\hat C_{2}[{\rm SU(3)}] + 2\hat{N} (2\hat{N} +3)$
is related to the quadratic Casimir operator of SU(3),
hence can be assigned to the DS-Hamiltonian, $\hat{H}_{\rm SU(3)-DS}$,
of Eq.~(\ref{SU3-ds}). In general,
$\hat{H}_{\rm PDS}$ of Eq.~(\ref{HSU3-PDS}) has SU(3)-PDS
with solvable ground $g(K=0)$ and $\gamma^{k}(K=2k)$ bands of
good SU(3) symmetry, while other bands,
in particular the $\beta(K=0)$ band, are mixed. 

The experimental spectra
of the  ground $g(K=0)$, $\gamma(K=2)$ and $\beta(K=0)$ bands
in $^{168}$Er
is shown in Fig.~\ref{figEr168}, and compared with an exact DS ($\eta_0=0$)
and PDS ($\eta_0\neq 0$) calculations~\cite{lev96}.
The SU(3) PDS spectrum is clearly seen to be an improvement over the
exact SU(3) DS
description, since the $\beta$-$\gamma$ degeneracy is lifted.
The ground and gamma are still pure SU(3) bands, but the beta
band is found to contain $13\%$ admixtures into the dominant $(2N-4,2)$
irrep~\cite{LevSin99}.
Since the wave functions of the solvable states are known, it is 
possible to obtain {\it analytic} expressions for matrix 
elements of observables between them.
The $E2$ operator can be transcribed as
$\hat{T}(E2) =
\alpha\, Q^{(2)} + \theta\, (d^{\dag}s + s^{\dag}\tilde{d})$,
with $Q^{(2)}$ an SU(3) generator.
Since the solvable ground and gamma bands reside in different SU(3) irreps, 
$Q^{(2)}$ cannot connect them and, consequently,
$B(E2)$ ratios for $\gamma\to g$ transitions do no depend on the $E2$
parameters ($\alpha,\theta$) nor on parameters of the
PDS Hamiltonian~(\ref{HSU3-PDS}).
Overall, as shown in the right panel of Fig.~1, these
parameter-free predictions of SU(3)-PDS
account well for the data in $^{168}$Er. Similar evidence for
SU(3)-PDS has been presented by Casten {\it et. al.} for other rare-earth
and actinide nuclei~\cite{Casten14,Couture15},
suggesting a wider applicability of this concept.
\begin{figure}[t]
  \centerline{
    \includegraphics[width=0.475\linewidth]{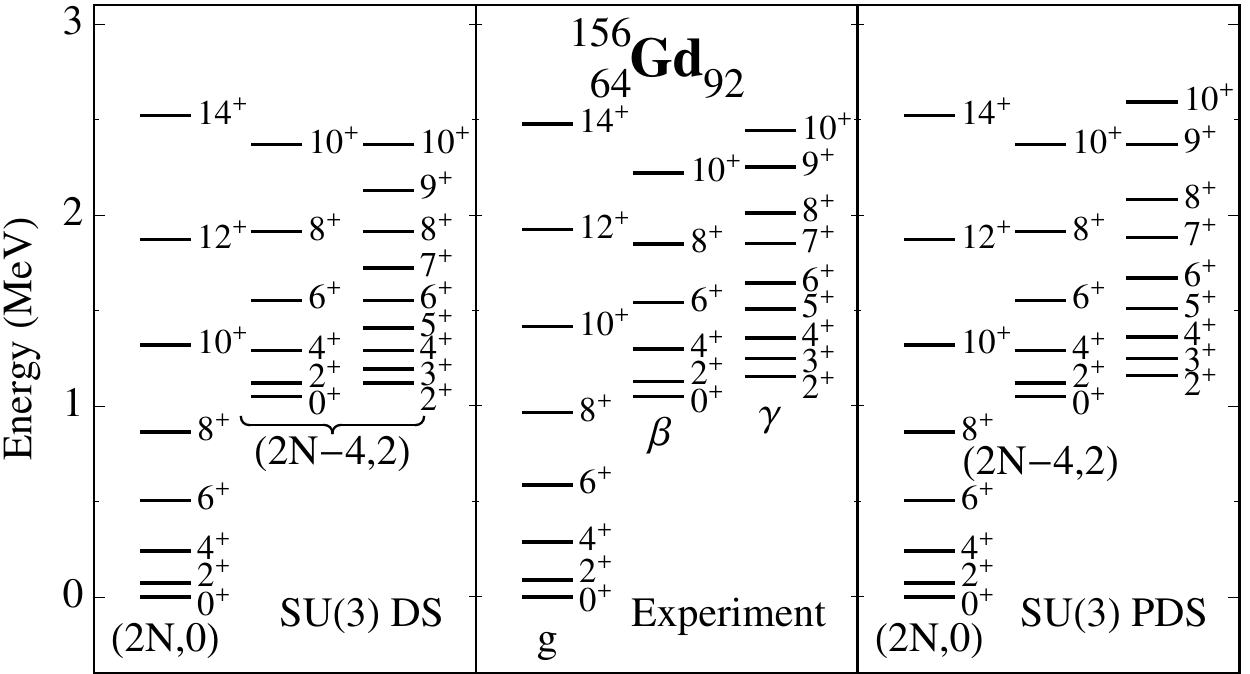}\hspace{0.2cm}%
    \includegraphics[width=0.5\linewidth]{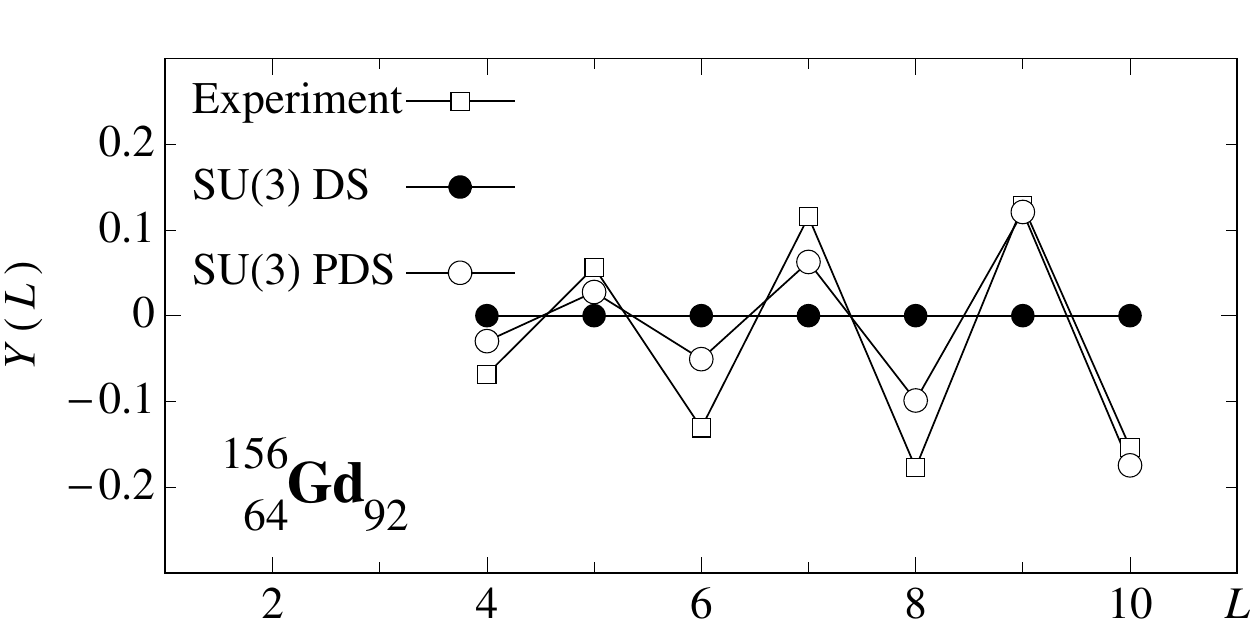}}
  \caption{
\small
Observed spectrum of $^{156}$Gd (left panels) and
odd-even staggering of the $\gamma$ band (right panel),
compared with SU(3)-DS and SU(3)-PDS calculations. The latter employs
$\hat{H}_{\rm PDS}$ of Eq.~(\ref{HSU3-PDS2}) with
$A\!=\!-7.6,\,C\!=\!12,\,\eta_2\!=\!-18.1,\,\eta_3\!=\!46.2$ keV
and $N\!=\!12$.
Here
$Y(L)=\frac{2L-1}{L}\times\frac{E(L)-E(L-1)}{E(L)-E(L-2)}-1$,
where $E(L)$ is the energy
of a $\gamma$-band level with angular momentum $L$.
Adapted from~\cite{Leviatan13}.}
\label{figGd156}
\end{figure}

Another class of Hamiltonians with SU(3) PDS exists, constructed 
of the following three-boson operators,
\ba
W^{\dag}_{3 m} &=& (P^{\dag}_{2}\,d^{\dag})^{(3)}_{m} =
\sqrt{7}\,[(d^{\dag} d^{\dag})^{(2)}d^{\dag}]^{(3)}_{m} ~,
\label{W3}
  \\ 
W^{\dag}_{4 m} &=& (P^{\dag}_{2}\,d^{\dag})^{(4)}_{m} ~,
\label{W4}
\ea
where $P^{\dag}_{2m}$ is given in Eq.~(\ref{P2}).
The $W^{\dag}_{\ell m}$ operators
are $(2,2)$ tensors under SU(3) and satisfy,
\ba
W_{\ell m}\,\ket{N,\, (\lambda,\mu)\!=\!(2N,0),\,K\!=\!0,\, L} &=& 0 ~,
\qquad\; \ell=2,3
\nonumber\\
W_{\ell m}\,
\ket{N,\, (\lambda,\mu)\!=\!(2N-4,2),\,K\!=\!0,\, L} &=& 0 ~.
\label{WL}
\ea
The above indicated states span the irrep $(2N,0)$ and part of the irrep
$(2N-4,2)$ of SU(3). The resulting SU(3)-PDS Hamiltonian has the
form~\cite{{Leviatan13}},
\ba
\hat{H}_{\rm PDS} =
\hat{H}_{\rm SU(3)-DS}+
\eta_2\,W_2^\dag\!\cdot\!\tilde{W}_2
+ \eta_3\,W_3^\dag\!\cdot\!\tilde{W}_3 ~,
\label{HSU3-PDS2}
\ea
where $\hat{H}_{\rm SU(3)-DS}$ is the SU(3)-DS Hamiltonian of
Eq.~(\ref{SU3-ds}).
Relations (\ref{WL}) ensure that $\hat{H}_{\rm PDS}$ has solvable
ground and $\beta$ bands with good SU(3) symmetry,
while other bands, in particular the $\gamma$ band,
are mixed.

A comparison of the experimental spectrum of $^{156}$Gd
with the SU(3)-DS calculation in Fig.~\ref{figGd156},
shows a good description for properties of 
states in the ground and $\beta$ bands,
however, the resulting fit to energies of the $\gamma$-band
is quite poor. The latter are not degenerate with the $\beta$ band and,
moreover, display an odd-even staggering with pronounced deviations 
from a rigid-rotor $L(L+1)$ pattern.
This effect can be visualized by plotting the quantity
$Y(L)$, defined in the caption of Fig.~\ref{figGd156}.
For a rotor this quantity is flat, $Y(L)=0$,
as illustrated in the right panel of 
Fig.~\ref{figGd156} with the SU(3) DS calculation, 
which is in marked disagreement with the empirical data. 
In the PDS calculation, the gamma band contains
$15\%$ SU(3) admixtures into the dominant $(2N-4,2)$ irrep 
and the empirical odd-even staggering is well reproduced.
The PDS results for the $\gamma$ band are obtained without 
affecting the solvability and SU(3) purity of states 
in the ground and beta bands. Since for the latter states the
wave functions are known, one has at hand closed expressions for
$E2$ transitions between them that can be used as tests for SU(3)-PDS.
Recent measurements by Aprahamian {\it et. al.} of 
lifetimes for $E2$ decays from states of the $\beta$-band 
in $^{156}$Gd~\cite{Aprahamian18},
confirm the PDS predictions.
\begin{figure}[t]
  \centerline{
    \includegraphics[width=0.77\linewidth]{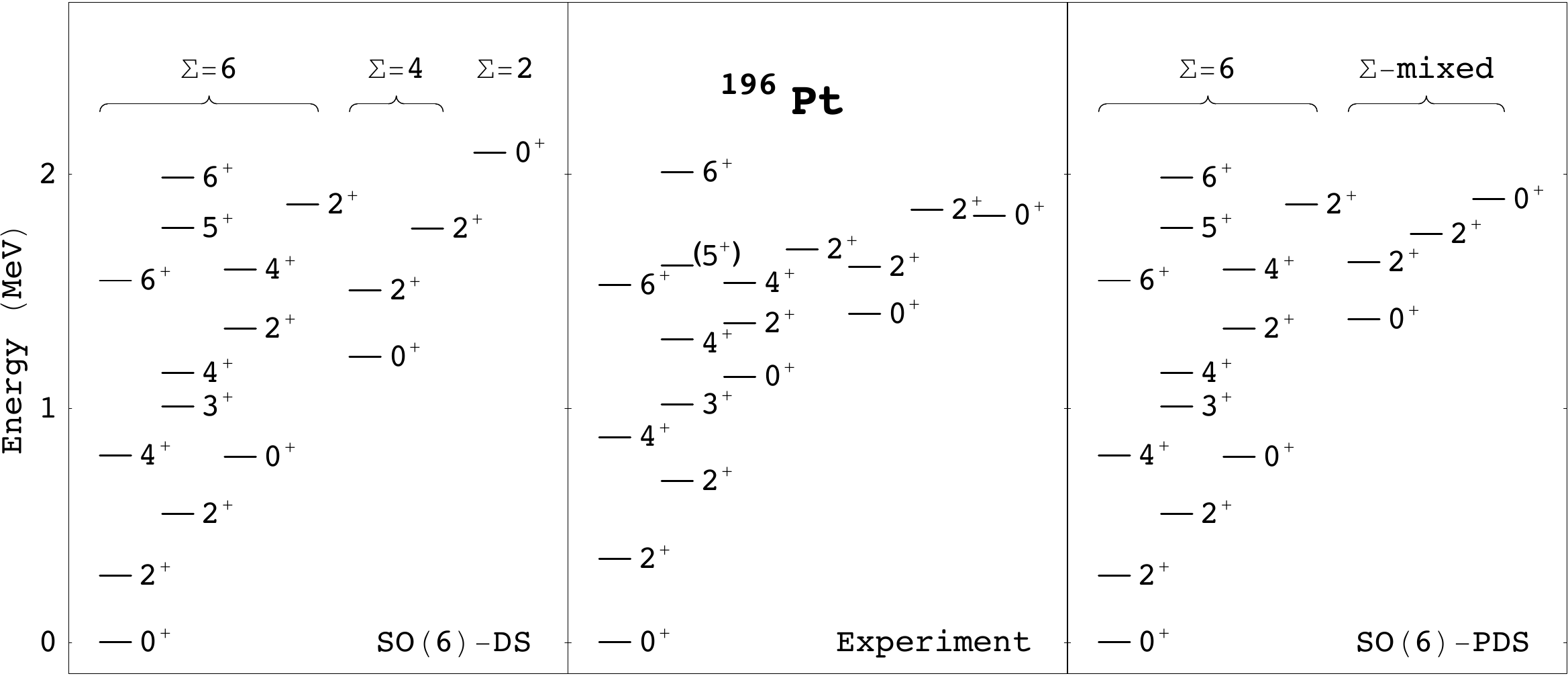}}
  \caption{
\small
Observed spectrum of $^{196}$Pt
compared with SO(6)-DS and SO(6)-PDS calculations. The latter employs
$\hat{H}_{\rm PDS}$ of Eq.~(\ref{HSO6-PDS}) with
$r_0\!=\!16.41,\,r_2\!=\!7.68,\,B\!=\!44.0,\, C\!=\!17.9$ keV
and $N\!=\!6$. $\Sigma$ is an SO(6) label.
Adapted from~\cite{RamLevVan09}.}
\label{pt196}
\end{figure}

The spectrum corresponding to the SO(6)-DS chain of Eq.~(\ref{SO6-ds}),
resembles that of a $\gamma$-unstable deformed 
roto-vibrator, where states are arranged 
in SO(6) multiplets with $\sigma=N-2v$ $(v=0,1,2,\ldots)$, 
and exhibit $\tau(\tau+3)$ and $L(L+1)$ rotational splitting.
To construct Hamiltonians with SO(6)-PDS, we consider
the three-boson operators
$s^{\dag}R^{\dag}_0$ and $d^{\dag}_{m}R^{\dag}_0$, 
where $R^{\dagger}_{0} = d^{\dagger}\cdot d^{\dagger} - (s^{\dagger})^2$.
These $\sigma=1$ tensors of SO(6)
annihilate all $(\tau,n_{\Delta},L)$ states in the irrep $\sigma=N$,
\ba
sR_{0}\,\vert N,\sigma=N,\tau, n_{\Delta}, L\rangle &=& 0 ~,
\qquad\quad \tau=0,1,2,\ldots,N
\nonumber\\
d_{m}R_{0}\,\vert N,\sigma=N,\tau, n_{\Delta}, L\rangle &=& 0 ~.
\label{R0vanish}
\ea
The intrinsic Hamiltonian reads
$\hat{H} = r_0\,R^{\dag}_0\hat{n}_sR_0 + r_2\,R^{\dag}_0\hat{n}_dR_0$ and
the  collective Hamiltonian is composed of
the quadratic Casimir operators of SO(5) and SO(3).
The complete Hamiltonian with SO(6)-PDS has the form~\cite{RamLevVan09},
\ba
\hat{H}_{\rm PDS} =
r_0\,R^{\dag}_0\hat{n}_sR_0 + r_2\,R^{\dag}_0\hat{n}_dR_0
+ B\,\hat{C}_{2}[\rm SO(5)] + C\,\hat{C}_{2}[\rm SO(3)]
= \hat{H}_{\rm SO(6)-DS} + \eta\,R^{\dag}_0\hat{n}_sR_0 ~,
\label{HSO6-PDS}
\ea
where the last equality follows from the fact that for
$r_2\!=\!r_0$, the intrinsic Hamiltonian
is related to the Casimir operator of SO(6),
$R^{\dag}_0R_0 \!=\! -\hat C_{2}[{\rm SO(6)}] + \hat{N} (\hat{N}+4)$,
hence can be assigned to the DS-Hamiltonian, $\hat{H}_{\rm SO(6)-DS}$,
of Eq.~(\ref{SO6-ds}).

A comparison with the experimental spectrum of
$^{196}$Pt in Fig.~3 and available $E2$ rates,  
reveals that the SO(6)-DS limit provides a good description for properties 
of states in the ground band.
However, the resulting fit to energies of excited bands is quite poor.
The $0^+_1$, $0^+_3$, and $0^+_4$ levels of $^{196}$Pt
at excitation energies 0, 1403, 1823 keV, respectively,
are identified as the bandhead states
of the ground $(v\!=\!0)$, first- $(v\!=\!1)$
and second- $(v\!=\!2)$ excited $\beta$-vibrational bands.
Their empirical anharmonicity,
defined by the ratio $R=E(v=2)/E(v=1)-2$,
is found to be $R=-0.70$.
The SO(6)-DS value is $R=-0.29$,
in marked disagreement with the empirical value. For the 
SO(6)-PDS Hamiltonian~(\ref{HSO6-PDS}), the ground band remains 
solvable with good SO(6) symmetry ($\sigma\!=\!N)$,
while the excited bands exhibit strong SO(6) breaking.
The calculated PDS anharmonicity
is $R=-0.63$, much closer to the empirical value~\cite{RamLevVan09}.
\begin{figure}[t]
  \centerline{
\fbox{\includegraphics[width=0.96\linewidth]{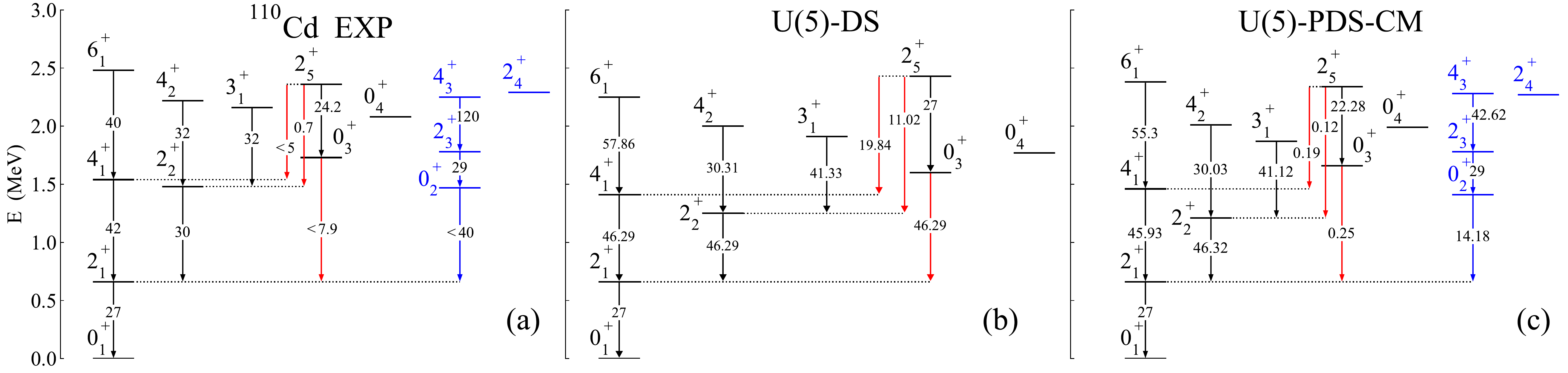}}}
\caption{
\small
Experimental spectrum and representative $E2$ rates
(in W.u.)  of normal and intruder levels
($0^{+}_2,\,2^{+}_3,\,4^{+}_3,\,2^{+}_4$)
in $^{110}$Cd, compared with U(5)-DS and U(5)-PDS calculations.
The latter employs
$\hat{H}_{\rm PDS}$ of Eq.~(\ref{HU5-PDS}) with parameters 
$\epsilon_d\!=\!473.35,\, A\!=\!-B\!=\!73.62,\, 
C\!=\!18.47,\,r_0\!=\!2.15,\, e_0\!=\!-6.92$
keV and $N\!=\!7$ in the normal sector.
The complete Hamiltonian contains additional interactions acting
in the intruder sector and a small configuration-mixing (CM) term,
see Ref.~\cite{GavLevRamVan18} for more details.}
\label{fig-cd110}
\end{figure}

The spectrum corresponding to the U(5)-DS chain of Eq.~(\ref{U5-ds}),
resembles that of a spherical vibrator, where states are arranged 
in U(5) multiplets: ($n_d\!=\!L\!=\!0$), ($n_d\!=\!1,L\!=\!2$), 
$(n_d\!=\!2,L\!=\!4,2,0$) and ($n_d\!=\!3,L\!=\!6,4,3,0,2$), 
with strong connecting $n_d+1\to n_d$ $E2$ transitions.
The U(5) basis states 
$\ket{N,\, n_d,\,\tau,\,n_{\Delta},\,L}$ have definite $d$-boson 
number $n_d$ and seniority $\tau$, and  $n_{\Delta}$ counts the maximum 
number of $d$-boson triplets coupled to $L\!=\!0$.
The construction of Hamiltonians with U(5)-PDS follows the general 
algorithm. The three-boson operators,
$G^{\dag}_{0} \!=\! [(d^\dag d^\dag)^{(2)} d^\dag]^{(0)}$ and
$K^{\dag}_{0} \!=\! s^{\dag}(d^{\dag} d^{\dag})^{(0)}$
are, respectively, $n_d=3$ and $n_d=2$ tensors with respect to U(5), 
and satisfy
\ba
G_0\ket{ N, n_d=\tau, \tau, n_{\Delta}=0, L} &=& 0 ~,
\qquad\quad L\!=\!\tau,\tau+1,\ldots,2\tau-2,2\tau 
\nonumber\\
K_0\ket{ N, n_d=\tau, \tau, n_{\Delta}=0, L} &=& 0 ~.
\label{V0vanish}
\ea
The resulting U(5)-PDS Hamiltonian contains the U(5)-DS
Hamiltonian~(\ref{U5-ds}) and terms constructed of
$G_0$ and $K_0$~\cite{GavLevRamVan18}, 
\ba
\hat{H}_{\rm PDS} =
\hat{H}_{\rm U(5)-DS} + r_0\,G^{\dag}_{0}G_{0}
+ e_{0}\,\left (G^{\dag}_0 K_0 + K^{\dag}_{0}G_0 \right ) ~.
\label{HU5-PDS}
\ea
$\hat{H}_{\rm PDS}$ has the U(5) basis states of Eq.~(\ref{V0vanish})
as eigenstates, with energies as in Eq.~(\ref{U5-ds}), while other
states are mixed.

The empirical spectrum of $^{110}$Cd, shown in Fig.~4(a), 
consists of both normal and intruder levels, the latter 
based on 2p-4h proton excitations across the $Z\!=\!50$ closed shell.
A comparison of the calculated spectrum [Fig.~4((b)] 
and $B(E2)$ values obtained from the DS limit~(\ref{U5-ds}),
demonstrates that most normal states have good spherical-vibrator
properties and conform well with the properties of U(5)-DS. 
However, the measured 
rates for $E2$ decays from the non-yrast states, $0^{+}_3\,(n_d\!=\!2)$ 
and $[0^{+}_4,\, 2^{+}_5\,(n_d\!=\!3)]$, 
reveal marked deviations from this behavior. 
In particular, 
$B(E2;\,0^{+}_3\!\rightarrow\! 2^{+}_1) \!<\! 7.9$, 
$B(E2;\,2^{+}_5\!\rightarrow\! 4^{+}_1) \!<\! 5$,
$B(E2;\,2^{+}_5\!\rightarrow\! 2^{+}_2) \!=\! 0.7^{+0.5}_{-0.6}$ 
Weisskopf units (W.u.), are extremely small compared to the 
U(5)-DS values: $46.29$, $19.84$, $11.02$ W.u., respectively. 
In a recent work~\cite{GavLevRamVan18}, $
\hat{H}_{\rm PDS}$ of Eq.~(\ref{HU5-PDS})
was taken to be the Hamiltonian in the normal sector
with a small mixing to the intruder sector, in the framework 
of the interacting boson model with configuration mixing (IBM-CM).
The resulting spectra, shown in Fig.~4(c), provides a good description 
of the empirical data in $^{110}$Cd. 
The majority of normal yrast states in the spectrum, 
correspond to the states of Eq.~(\ref{V0vanish}), 
and maintain the good U(5) symmetry, to a good approximation.
In contrast, the U(5) structure of the non-yrast states changes dramatically.
The resulting calculated values:
$B(E2;\,0^{+}_3\!\rightarrow\! 2^{+}_1) \!=\! 0.25$, 
$B(E2;\,2^{+}_5\!\rightarrow\! 4^{+}_1) \!=\! 0.19$ and 
$B(E2;\,2^{+}_5\!\rightarrow\! 2^{+}_2) \!=\! 0.12$ W.u.,
are consistent with the measured values~\cite{GavLevRamVan18}.

\section{MULTIPLE PARTIAL DYNAMICAL SYMMETRIES AND SHAPE COEXISTENCE}

Symmetry plays a profound role in quantum phase transitions (QPTs), 
which are qualitative changes in the structure
of a physical system, induced by a variation of a coupling constant in the
quantum Hamiltonian. Such ground state phase transitions~\cite{Gilmore78},
are a pervasive phenomenon observed in many branches of physics~\cite{carr},
and are  realized empirically in nuclei as transitions between different 
shapes~\cite{CejJolCas10,iachello11,IacLevPet11,PetLevIac11}. 
QPTs occur as a result of competing terms in the Hamiltonian with
incompatible (non-commuting) symmetries.
An interesting question to address is 
whether there are any symmetries (or traces of) still present 
in such circumstances, especially at the critical point where the
structure changes most rapidly and mixing effects are enhanced. 
The feasibility of such persisting symmetries gained support
from the works by Iachello 
on ``critical-point symmetries''~\cite{E5,X5},
which demonstrated that the dynamics in such environment
is amenable to {\it analytic} descriptions.
A convenient framework to study symmetry-aspects of QPT in nuclei 
is the IBM~\cite{ibm}, 
whose dynamical symmetries, Eqs.~(3)-(6),
correspond to possible phases of the system.
The relevant Hamiltonians mix terms from different DS chains,
$\hat{H}(\xi)\propto (1-\xi)\hat{H}_1 + \xi\,\hat{H}_2$.
The nature of the phase transition is governed by the topology of 
the corresponding surface~(\ref{enesurf}),
which serves as a Landau's potential,
with the equilibrium deformations as order parameters,
and the coupling constant $\xi$ as a control parameter.
In the present contribution,
we focus on remaining symmetries at the critical points of first-order
QPTs, where the underlying surface exhibits multiple degenerate minima,
with different types of dynamics (and symmetry) associated with each
minimum. In such circumstances, exact DSs
are broken and surviving symmetries, if any, are at most partial.
\begin{figure}[t]
  \centering
\includegraphics[width=16cm]{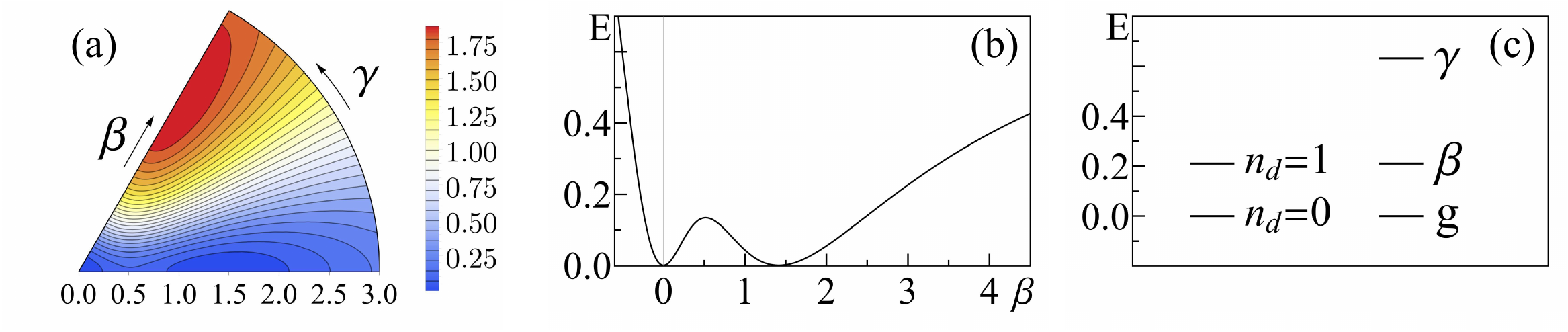}
  \caption{
Spherical-prolate (S-P) shape coexistence.
(a)~Contour plots of the energy surface~(\ref{enesurf}),   
(b)~$\gamma\!=\!0$ sections, and
(c)~bandhead spectrum, for the Hamiltonian $\hat{H}_{\rm PDS}$
of Eq.~(\ref{Hu5su3}), with multiple U(5)-PDS and SU(3)-PDS.
Adapted from~\cite{LevGav17}.}
\label{fig5-SP}
\end{figure}

The construction of Hamiltonians with several distinct partial dynamical 
symmetries, requires an extension of the previously discussed PDS algorithm.
For that purpose, consider two DS chains of the IBM,
\ba
{\rm U(6)\supset G_1\supset G_2\supset \ldots \supset SO(3)} &&\;\;\quad
\ket{N,\, \lambda_1,\,\lambda_2,\,\ldots,\,L} ~,
\label{ds-G1}\\
{\rm U(6)\supset G'_1\supset G'_2\supset \ldots \supset SO(3)} &&\;\;\quad
\ket{N,\, \sigma_1,\,\sigma_2,\,\ldots,\,L} ~, 
\label{ds-G1prime}
\ea
with different leading sub-algebras ($G_1\!\neq\! G'_1$) and associated bases.
We seek $n$-particle annihilation
operators $\hat{T}_{\alpha}$ which satisfy simultaneously the following
two conditions,
\ba
\hat{T}_{\alpha}
\ket{N,\lambda_1\!=\Lambda_0,\lambda_2,\ldots,L} 
&=& 0 ~,
\label{basis1}\\
\hat{T}_{\alpha}
\ket{N,\sigma_1\!=\!\Sigma_0,\sigma_2,\ldots,L} 
&=&0 ~.
\label{basis2}
\ea
The states of Eq.~(\ref{basis1}) reside in the $\lambda_1=\Lambda_0$ irrep 
of $G_1$, are classified according to the DS-chain (\ref{ds-G1}), hence 
have good $G_1$ symmetry. Similarly,  
the states of Eq.~(\ref{basis2}) reside in the $\sigma_1=\Sigma_0$ irrep 
of $G'_1$, are classified according to the DS-chain (\ref{ds-G1prime}), 
hence have good $G'_1$ symmetry.
The intrinsic Hamiltonian, $\hat{H}$, has the same form as
in Eq.~(\ref{H-int}).
Although $G_1$ and $G'_1$ are incompatible, relations
(\ref{basis1})-(\ref{basis2}) ensure that
both sets of states are eigenstates of the same Hamiltonian.
In general, $\hat{H}$ itself is not necessarily 
invariant under $G_1$ nor under $G_2$ and, therefore, its other eigenstates 
can be mixed with respect to both $G_1$ and $G'_1$. 
The collective Hamiltonian, $\hat{H}_c$, has the form as in Eq.~(\ref{H-col}), 
but now includes the Casimir operators of algebras
which are common to both chains, and generate rotational splitting.
The resulting complete Hamiltonian,
$\hat{H}_{\rm PDS} = \hat{H} + \hat{H}_c$, Eq.~(\ref{H-PDS}),
has both $G_1$-PDS and $G'_1$-PDS. 
The case of triple (or multiple) PDSs,
associated with three (or more) incompatible DS-chains,
is treated in a similar fashion. 

In conjunction with quantum phase transitions, 
the two DS chains of Eqs.~(\ref{ds-G1})-(\ref{ds-G1prime})
describe the dynamics of different shapes, specified by
equilibrium deformations ($\beta_1,\gamma_1$) and ($\beta_2,\gamma_2$). 
The derived PDS Hamiltonian has a potential surface with
two degenerate minima, hence is qualified as a critical-point Hamiltonian.
The two sets of solvable eigenstates, Eqs.~(\ref{basis1})-(\ref{basis2}),
span the ground bands of the two shapes associated with the two minima.
In what follows, we apply the above procedure 
to a variety of coexisting shapes and related multiple-PDSs
in the IBM~framework. 

The U(5)-DS and SU(3)-DS chains, Eqs.~(\ref{U5-ds}) and~(\ref{SU3-ds}),
are relevant to spherical and prolate-deformed shapes.
The construction of PDS Hamiltonian suitable for the coexistence
of such shapes, follows the above procedure. The two-boson operator
$P_{2m}$ of Eq.~(\ref{P2}) annihilates the states of the
SU(3) irrep $(2N,0)$ and the U(5) irrep, $n_d=0$,
\ba
P_{2m}\,\ket{N,\, (\lambda,\mu)\!=\!(2N,0),\,K\!=\!0,\, L} &=& 0 
\qquad\qquad L=0,2,4,\ldots,2N
\label{2N0}\\
P_{2m}\,\ket{N,\, n_d=0,\,\tau=0,\,L=0} &=& 0 ~,
\label{nd0}
\ea
and corresponds to the $\hat{T}_{\alpha}$ of
Eqs.~(\ref{basis1})-(\ref{basis2}).
The resulting PDS Hamiltonian is given by~\cite{lev07},
\ba
\hat{H}_{\rm PDS} = h_{2}\,P^{\dagger}_{2}\cdot \tilde{P}_{2}
+ C\,\hat{C}_{2}[\rm SO(3)] ~,
\label{Hu5su3}
\ea
and has a solvable prolate-deformed ground band
with good SU(3) symmetry and a solvable spherical $L=0$ ground state
with good U(5) symmetry.
$\hat{H}_{\rm PDS}$ has additional solvable SU(3) basis states
$\ket{N,(\lambda,\mu)\!=\!(2N-4k,2k)K\!=\!2k,L}$, which span the deformed 
$\gamma^k(K\!=\!2k)$ bands, and an additional solvable U(5) basis state 
with $\ket{N,n_d\!=\!\tau\!=\!L\!=\!3}$. 
Other eigenstates are mixed with respect to both U(5) and SU(3).
Altogether, $\hat{H}_{\rm PDS}$ has U(5)-PDS coexisting with SU(3)-PDS.
The corresponding energy surface, shown in Fig.~5,
has degenerate spherical ($\beta_{\rm eq}\!=\!0$) and prolate-deformed
($\beta_{\rm eq}\!=\!\sqrt{2},\gamma_{\rm eq}\!=\!0$) minima, separated
by a barrier. $\hat{H}_{\rm PDS}$ thus qualifies as critical-point
Hamiltonian. The normal modes involve $\beta$ and $\gamma$ vibrations
about the deformed minimum and quadrupole vibrations about the
spherical minimum. The bandhead spectrum associated with these modes, 
is shown in Fig.~5(c).
\begin{figure}[t]
\centering
\includegraphics[width=16cm]{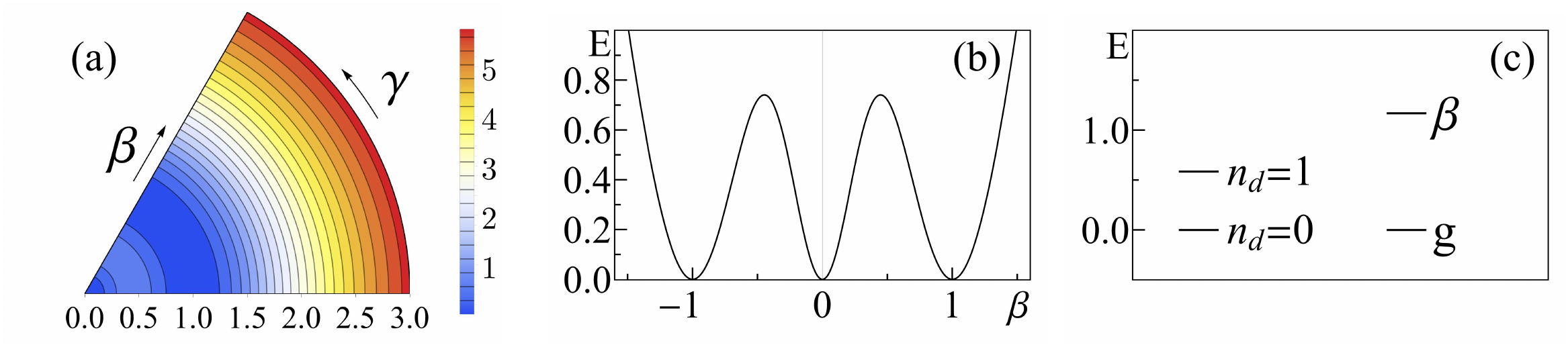}
\caption{\label{fig6-SG}
\small
Spherical and $\gamma$-unstable deformed (S-G) shape coexistence.
(a)~Contour plots of the energy surface~(\ref{enesurf}),
(b)~$\gamma\!=\!0$ sections, and
(c)~bandhead spectrum, for
$\hat{H}_{\rm PDS}$ of Eq.~(\ref{Hu5so6}),
with multiple U(5)-PDS and SO(6)-PDS.
Adapted from~\cite{LevGav17}.}
\end{figure}

The U(5)-DS and SO(6)-DS chains, Eqs.~(\ref{U5-ds}) and~(\ref{SO6-ds}),
are relevant to spherical and $\gamma$-unstable deformed shapes.
To construct a PDS Hamiltonian suitable for the coexistence of these
shapes, we consider the operator $d_{m}R_{0}$ of Eq.~(\ref{R0vanish}), 
which annihilates the $(\tau,n_{\Delta},L)$ states of the
SO(6) irrep $\sigma=N$ and the single $L=0$ state of the
U(5) irrep, $n_d=0$,
\ba
d_{m}R_0\,
\ket{N,\, \sigma=N,\,\tau,\, n_{\Delta},\,L} &=& 0 
\qquad\qquad \tau=0,1,2,\ldots,N
\label{sigmaN}
\\
d_{m}R_0\,
\ket{N,\, n_d=0,\,\tau=0,\,L=0} &=& 0 ~.
\label{nd0SG}
\ea
The resulting PDS Hamiltonian is
given by~\cite{LevGav17},
\ba
\hat{H}_{\rm PDS} &=& r_2\,R^{\dag}_0\hat{n}_dR_0
+ B\,\hat{C}_2[\rm SO(5)] + C\,\hat{C}_2[\rm SO(3)] ~,
\label{Hu5so6}
\ea 
and has a solvable
$\gamma$-unstable deformed ground band
with good SO(6) symmetry and a solvable spherical ground state
with good U(5) symmetry.
Other eigenstates are mixed with respect to both U(5) and SO(6).
Altogether, $\hat{H}_{\rm PDS}$ has U(5)-PDS coexisting with SO(6)-PDS.
The corresponding energy surface, shown in Fig.~6,
is independent of $\gamma$, in accord with the SO(5) symmetry of the
Hamiltonian. It has two degenerate minima at $\beta_{\rm eq}\!=\!0$ and
($\beta_{\rm eq}\!=\!1,\gamma_{\rm eq}$ arbitrary), separated by a barrier.
$\hat{H}_{\rm PDS}$ thus qualifies as critical-point Hamiltonian.
The normal modes involve $\beta$ vibrations about the $\gamma$-unstable
deformed minimum and quadrupole vibrations about the spherical minimum.

The SU(3)-DS and $\bsu3$-DS, Eqs.~(\ref{SU3-ds}) and (\ref{SU3bar-ds}),
are relevant to prolate and oblate shapes, respectively.
The two chains have similar properties, but the classification 
of states is different.
The ground band spans the irrep $(\lambda,\mu)=(2N,0)$
[$(\blam,\bmu)=(0,2N)$], in SU(3)-DS [$\bsu3$-DS], 
while the $\beta$ and $\gamma$ bands 
reside in the irrep $(\lambda,\mu)=(2N-4,2)$ 
[$(\blam,\bmu)=(4,2N-4)$].
In Figs.~7-8, we denote such prolate and oblate bands by 
$(g_1,\beta_1,\gamma_1)$ and ($g_2,\beta_2,\gamma_2$), respectively.
To construct a PDS Hamiltonian appropriate for 
prolate-oblate coexistence, we consider the three-boson operators:
$s^{\dag}P^{\dag}_0,\,d^{\dag}_{m}P^{\dag}_0,\,W^{\dag}_{3m}$, which satisfy
\ba
& sP_0\,\ket{N,\,(\lambda,\mu)=(2N,0),\,K=0,\,L} = 0 \;, \;\; 
&sP_0\,\ket{N,\, (\blam,\bmu)=(0,2N),\,\bar{K}=0,\, L} = 0 ~,
\label{sp0}
\nonumber\\
& d_{m}P_0\,\ket{N,\,(\lambda,\mu)=(2N,0),\,K=0,\,L} = 0\; , \;\;
&d_{m}P_0\,\ket{N,\, (\blam,\bmu)=(0,2N),\,\bar{K}=0,\, L} = 0 ~, 
\label{dp0}
\nonumber\\
& W_{3m}\,\ket{N,\,(\lambda,\mu)=(2N,0),\,K=0,\,L} = 0 \; , \;
&W_{3m}\,\ket{N,\, (\blam,\bmu)=(0,2N),\,\bar{K}=0,\, L} = 0 ~.
\label{w3m} 
\ea
Here $P^{\dag}_0$ and $W^{\dag}_{3m}$ are given in Eqs.~(\ref{P0}) 
and \ref{W3}), respectively. The PDS Hamiltonian is
found to be~\cite{LevDek16},
\ba
\hat{H}_{\rm PDS} &=& 
h_0\,P^{\dag}_0\hat{n}_sP_0 + h_2\,P^{\dag}_0\hat{n}_dP_0 
+\eta_3\,W^{\dag}_3\cdot\tilde{W}_3
+ C\,\hat{C}_2[\rm SO(3)] + \delta\,\hat{C}_{2}[SU(3)] ~.
\label{Hsu3su3b}
\ea 
The last term, with infinitesimally small $\delta$, is needed to
avoid an undesired invariance of the Hamiltonian with respect
to a phase change of the $s$-boson~\cite{LevDek16}.
$\hat{H}_{\rm PDS}$
has solvable prolate and oblate ground bands with good 
SU(3) and $\bsu3$ symmetry, respectively. 
Other eigenstates are mixed with respect to both SU(3) and $\bsu3$.
Altogether, $\hat{H}_{\rm PDS}$ has SU(3)-PDS coexisting with 
$\bsu3$-PDS.
The corresponding energy surface, shown in Fig.~7, 
has two degenerate minima at 
$(\beta_{\rm eq}\!=\!\sqrt{2},\gamma_{\rm eq}\!=\!0)$ and 
$(\beta_{\rm eq}\!=\!\sqrt{2},\gamma_{\rm eq}\!=\!\pi/3)$, separated 
by a barrier. $\hat{H}_{\rm PDS}$ thus qualifies as a 
critical-point Hamiltonian.
The normal modes involve $\beta$ and $\gamma$ 
vibrations about the respective deformed minima.
\begin{figure}[t]
  \centering
\includegraphics[width=16cm]{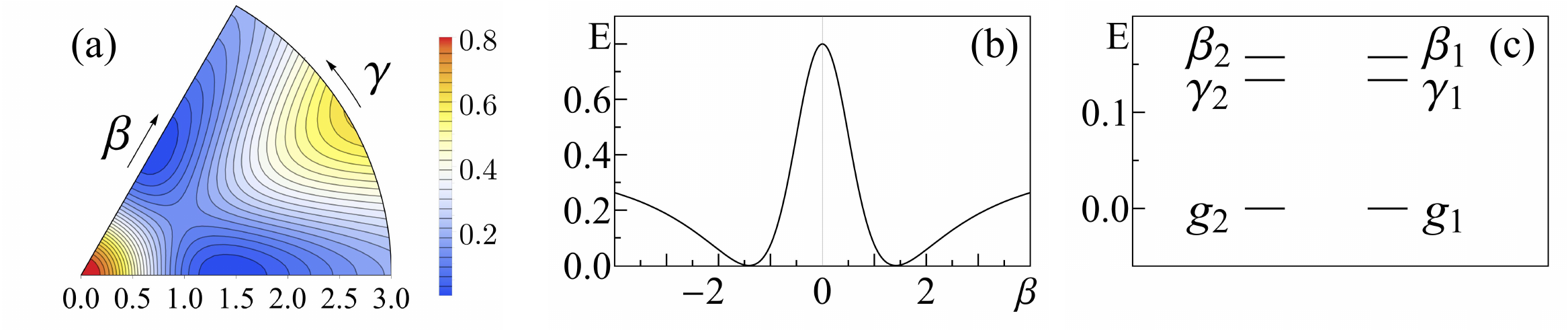}
\caption{\label{fig7-PO}
\small
Prolate-oblate (P-O) shape coexistence.
(a)~Contour plots of the energy surface~(\ref{enesurf}),   
(b)~$\gamma\!=\!0$ sections, and
(c)~bandhead spectrum, for
$\hat{H}_{\rm PDS}$ of Eq.~(\ref{Hsu3su3b}),
with multiple SU(3)-PDS and $\bsu3$-PDS.
Adapted from~\cite{LevGav17}.} 
\end{figure}
\begin{figure}[t]
\centering
\includegraphics[width=16cm]{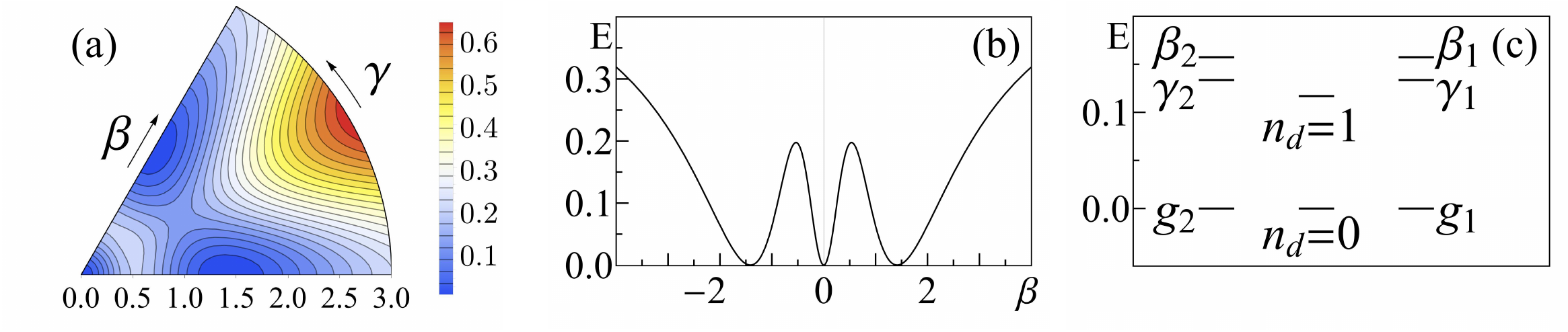}
\caption{\label{fig8-SPO}
\small
Spherical-prolate-oblate (S-P-O) shape coexistence.
(a)~Contour plots of the energy surface~(\ref{enesurf}),
(b)~$\gamma\!=\!0$ sections, and
(c)~bandhead spectrum, for
$\hat{H}_{\rm PDS}(h_0=0)$ of Eq.~(\ref{Hsu3su3b}),
with multiple U(5)-SU(3)-$\bsu3$ PDSs.
Adapted from~\cite{LevGav17}.} 
\end{figure}

For $h_0=0$, the Hamiltonian of Eq.~(\ref{Hsu3su3b}) 
exhibits multiple U(5), SU(3) and $\bsu3$ PDSs since
the $d_{m}P_0$ and $W_{3m}$ operators
of Eq.~(\ref{w3m}), annihilate also the following U(5) basis state,
\ba
d_{m}P_0\,
\ket{N,\, n_d=0,\,\tau=0,\,L=0} &=& 0 ~,
\label{dmp0-nd0}
\nonumber\\
W_{3m}\,
\ket{N,\, n_d=0,\,\tau=0,\,L=0} &=& 0 ~.
\label {w3-nd0}
\ea
Altogether, $\hat{H}_{\rm PDS}(h_0=0)$ has
a solvable spherical ground state with good U(5) symmetry,
in addition to solvable prolate and oblate
ground bands with good SU(3) and $\bsu3$ symmetry.
The corresponding energy surface, shown in Fig.~8, 
has three degenerate minima, at $\beta_{\rm eq}\!=\!0$ and
$(\beta_{\rm eq}\!=\!\sqrt{2},\gamma_{\rm eq}\!=\!0,\pi/3)$,
separated by barriers. 
$\hat{H}_{\rm PDS}(h_0=0)$ thus qualifies as a 
critical-point Hamiltonian for triple coexistence of 
spherical, prolate and oblate shapes.
In addition to $\beta$ and $\gamma$ vibrations,
the normal modes involve also quadrupole vibrations 
about the spherical minimum. 

In all cases of shape-coexistence and multiple-PDSs considered above,
one can obtain analytic expressions of quadrupole moments and transition
rates for the remaining solvable states,
which are the observables most closely related to the nuclear shape. 
These expressions can be used as signatures and tests for the underlying
PDSs. The purity and good quantum numbers of these selected states,
enable the derivation of symmetry-based selection rules for $E2$ and
$E0$ decays and, as shown Fig.~9, the subsequent identification of
isomeric states.

Partial dynamical symmetries have also been identified in coupled systems 
with $U_{1}(m)\otimes U_{2}(n)$ spectrum generating algebras. 
This includes, partial F-spin symmetry~\cite{Levgin00}
in the proton-neutron version of the interacting boson model
(IBM-2)~\cite{ibm} of even-even nuclei,
and  ${\rm SO_{B+F}(6)}$ partial Bose-Fermi symmetry~\cite{Pds-BF15} 
in the interacting boson-fermion model (IBFM)
of odd-mass nuclei~\cite{ibfm}. 
Hamiltonians with PDS are not completely regular nor fully chaotic.
As such they are relevant to the study of mixed systems with 
coexisting regularity and chaos~\cite{Whe93,LevWhe96,Macek14}.

\section{ACKNOWLEDGMENTS}
It is a pleasure and honor to dedicate this contribution to Francesco 
Iachello on the occasion of his retirement.
Franco: ``a one-man center of theoretical physics'', a rare combination of
true scholar, innovative scientist and friend.
I vividly recall and cherish many ``blackboard hours''
of in-depth discussions, thoughtful guidance, impact and inspiration.
Segments of the reported results were obtained in collaboration with 
N.~Gavrielov (HU), J.~E.~Garc\'\i a-Ramos (Huelva) and 
P.~Van~Isacker~(GANIL).
This work is supported by the Israel Science Foundation (Grant 586/16).
\begin{figure}[t]
  \centering
\includegraphics[width=0.36\linewidth]{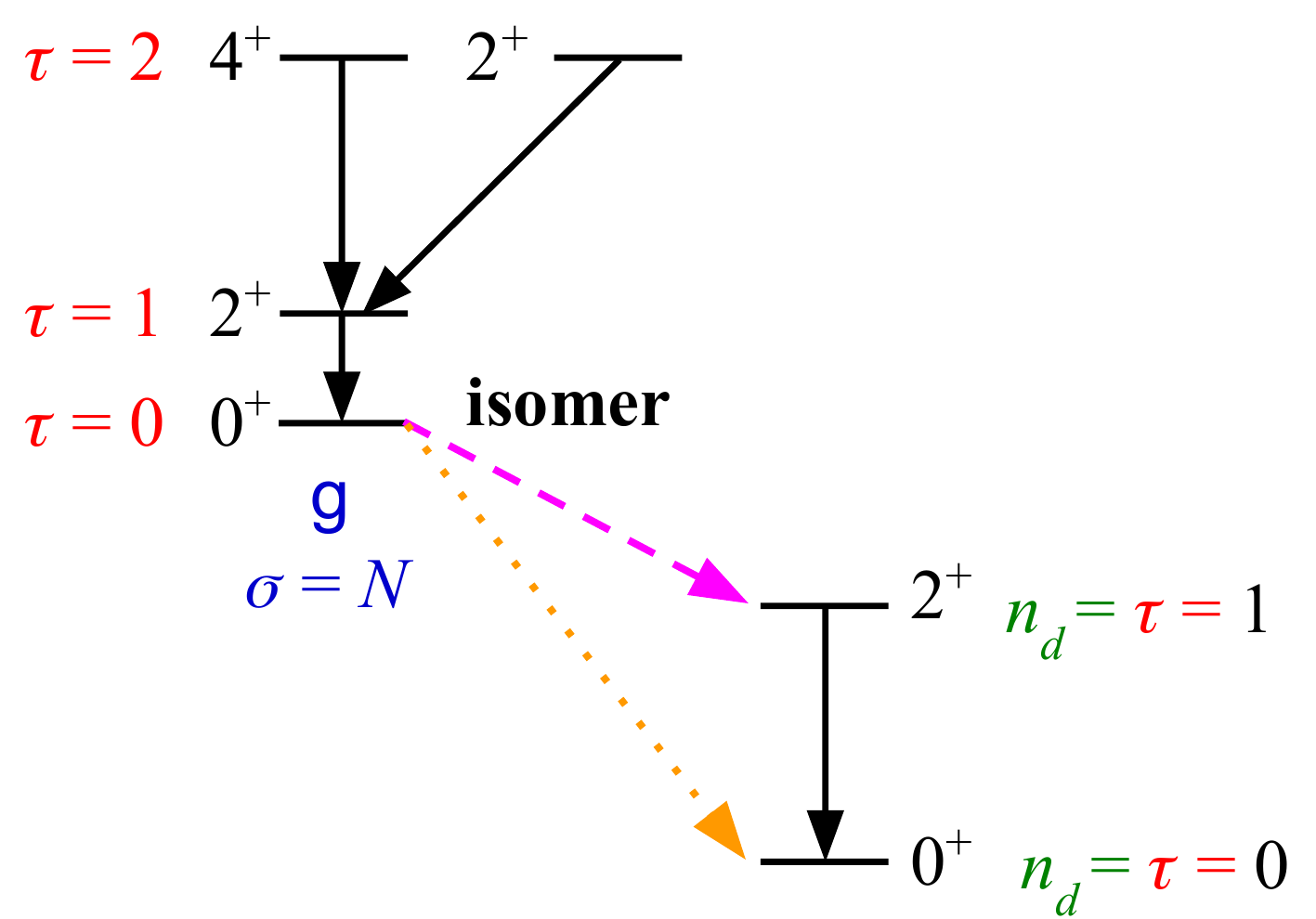}\hspace{-0.21cm}
\includegraphics[width=0.3\linewidth]{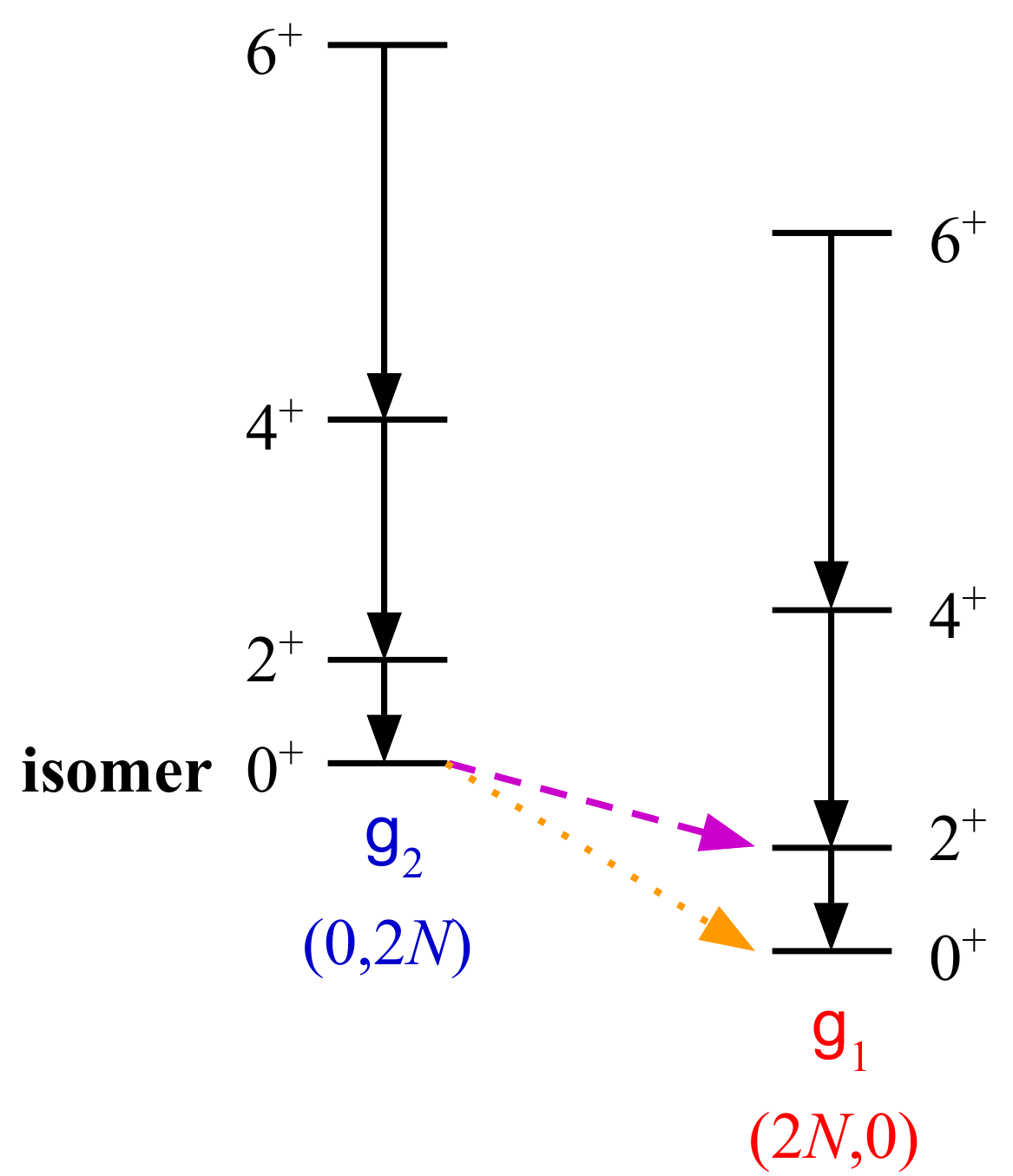}\hspace{0.3cm}
\includegraphics[width=0.31\linewidth]{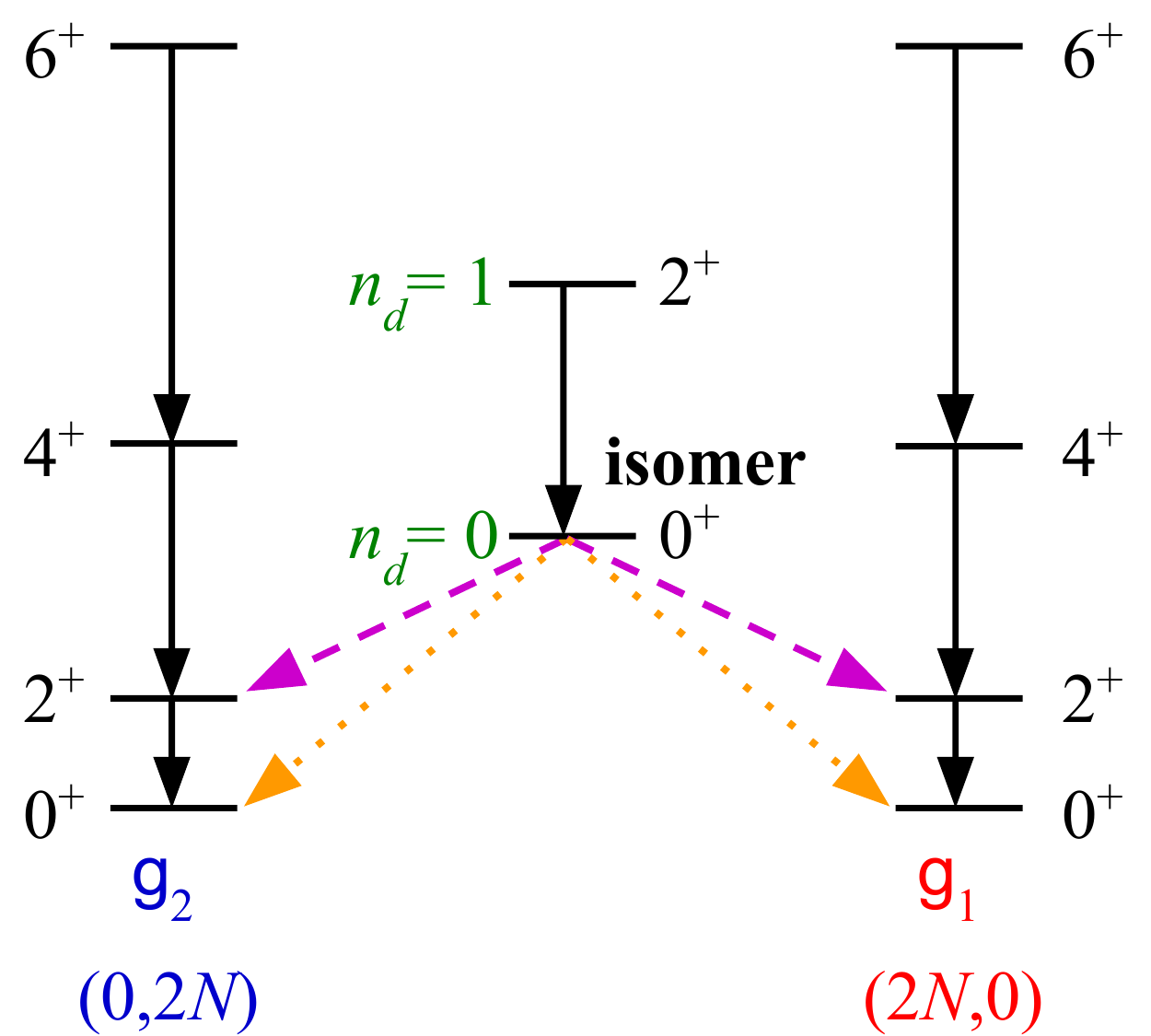}
\caption{\label{fig9SPObe2}
\small
Signatures of multiple U(5)-SO(6) PDSs (left panel),
SU(3)-$\bsu3$ PDSs (center panel),
and U(5)-SU(3)-$\bsu3$ PDSs (right panel),
relevant to S-G, P-O and S-P-O shape coexistence, respectively.
The rates for the strong intraband $E2$ transitions (solid lines) 
are known analytically.
Retarded $E2$ (dashes lines) and $E0$ (dotted lines) decays 
identify isomeric states.}
\end{figure}

\end{document}